\documentclass[prd]{revtex4}
\usepackage{array}
\usepackage{booktabs}
\usepackage{tabu}
\usepackage{dcolumn}
\usepackage{amsmath}
\usepackage{amsfonts}
\usepackage{amssymb}
\usepackage{graphicx}
\usepackage{subfigure}
\usepackage{graphicx}
\usepackage{dcolumn}
\usepackage{bm}

\def\be{\begin{equation}}
  \def\ee{\end{equation}}
\def\bea{\begin{eqnarray}}
\def\eea{\end{eqnarray}}
\def\f{\frac}
\def\n{\nonumber}
\def\l{\label}
\def\p{\phi}
\def\o{\over}
\def\R{\rho}
\def\pa{\partial}
\def\om{\omega}
\def\na{\nabla}
\def\P{\Phi}

\begin{document}

\title{Light of Planck-2015 on Non-Canonical Inflation}
\author{$^a$Kh. Saaidi}
 \email{ksaaidi@uok.ac.ir}
  \author{$^b$A. Mohammadi}
   \email{abolhassanm@gmail.com}
    \author{$^{a}$T. Golanbari}
     \email{t.golanbari@uok.ac.ir}

\affiliation{
$^a$Department of Physics, Faculty of Science, University of Kurdistan,  Sanandaj, Iran.
$^b$Young Researchers and Elites Club, Sanandaj Branch, Islamic Azad University, Sanandaj, Iran.\\}
\date{\today}

\def\be{\begin{equation}}
  \def\ee{\end{equation}}
\def\bea{\begin{eqnarray}}
\def\eea{\end{eqnarray}}
\def\f{\frac}
\def\n{\nonumber}
\def\l{\label}
\def\p{\phi}
\def\o{\over}
\def\R{\rho}
\def\pa{\partial}
\def\om{\omega}
\def\na{\nabla}
\def\P{\Phi}

\begin{abstract}
Slow-roll inflationary scenario is considered in non-canonical scalar field model supposing a power-law function for kinetic term, and using two formalisms. In the first approach, the potential is picked out as a power-law function, that is the most common approach in studying inflation. Hamilton-Jacobi approach is selected as the second formalism, so that the Hubble parameter is introduced as a function of scalar field instead of the potential. Employing the last observational data, the free parameters of the model are constrained, and the predicted form of the potential and attractor behavior of the model are studied in detail.
\end{abstract}
\maketitle

\section{Introduction}\label{Intro}
Inflationary scenario is the best candidate for describing the very early times evolution of the universe, which could simply solve the problems of the standard hot big bang model. There are lots of observational data supporting this scenario \cite{1}. The scenario was first proposed by American physicist, Alan Guth, in order to solve the problem of standard hot big bang model \cite{2}. So far, different model of inflationary cosmology have been introduced, however it might be said that the most interesting one is chaotic inflation, introduced by Linde in 1983 \cite{3}. Based on this scenario, the universe in the very early times is dominated by a slowly varying homogenous scalar field, and undergoes an exponentially expansion in very short time.\\
The scenario has been constructed based on canonical scalar field, however recently the cosmological models of scalar field including non-canonical kinetic term have attended scientists attention. The general form of its action is expressed by $\mathcal{L}_\phi = f(\phi)F(X)-V(\phi)$, where $X = (\nabla_\mu \phi \nabla^\mu \phi ) / 2$ \cite{4}. The case with $V(\phi)=0$ comes into a well-known model as k-essence. The main idea of k-essence comes from Born-Infold action of string theory \cite{5}. The model is able to give some interesting results about dark energy \cite{6,6a,6b,6c,6d,6e}. In \cite{7,7a}, the model is applied as a possible way for inflation and describing early time evolution of the universe. \\
In the present work, we are going to take $f(\phi)=1$, and $V(\phi)$ as scalar field potential. In other work, we take a pure kinetic k-essence plus a potential term. This kind of model is known as non-canonical scalar field \cite{8}. This case is another class of the general form which could be as important and interesting as k-essence model. The cosmological solution of the model has been studied in \cite{8}, where it was shown that it is possible to construct a unified model of dark matter and dark energy for a simple form of non-canonical kinetic term $F(X)$. The same case has been considered in \cite{9} as well, in which they found that producing a unified model of dark matter and dark energy for a pure kinetic k-essence is very difficult. It seems that the non-canonical scalar field model hass this ability to be an appropriate model of the universe evolution and has merit for considering in more detail. Then, we motivated to use non-canonical scalar field as a possible model for describing one of earliest universe evolution, namely inflation. \\
There are two formalisms that cosmologists apply to study inflation, and in the present work, we are going to utilized both these approach. The first formalism, and the most common and well-known one, relies on the potential. In this approach, we need to propose a specific form of the potential to be able to investigate the model in detail. In this regard, During this work, a familiar form of potential, as power-law, is picked out. The power-law potential, $\lambda \phi^\alpha$, has a huge contribution in inflationary scenario studies, which includes chaotic inflation. The two most known cases are $\alpha=4$ and $\alpha=2$. Recent observational results have shown that the $\lambda\phi^4$ stays outside of the joint $99.7\%$ CL region in the $n_s-r$ plane. The quadratic potential model, $\alpha=2$, lies outside the joint $95\%$ CL region for Planck+WP+high-\textit{l} for $N \leq 60$ e-folds. The potential models with $\alpha=1$ and $\alpha=2/3$ lie on the boundary of the joint $95\%$ CL region \cite{DataPot,DataPot-a,Planck1}. Another interesting formalism is known as Hamilton-Jacobi formalism, which relies on a specific form of the Hubble parameter in term of scalar field, instead of introducing a potential \cite{Baumann,HJ,HJ-a,HJ-b,HJ-c}. In comparison with the first formalism, Hamilton-Jacobi formalism includes some fascinating feature such as deducing a form of the potential, obtaining an exact solution for the scalar field. Although the formalism has got less attention than first formalism, its abilities in inflationary studies sounds to be undeniable. Based on the presented argument, the later formalism stands in the center of our attention in this work. \\
The main goal of the work is to constrain the free parameters of the model utilizing the latest observational data released by Planck 2015 \cite{Planck1,Planck2}. Prediction of quantum perturbation is one of interesting feature of inflationary scenario. The most important of these perturbations are scalar perturbation, which is seeds for large scale structure of the universe, and tensor perturbation, which is known as gravitational waves too \cite{Perturbation,Perturbation-a,Perturbation-b}. The prediction is supported by huge amount of observational data. From Planck data, the amplitude of scalar perturbation is about $\ln\Big( 10^{10}A^2_s \Big)=3.094$, and the scalar spectra index, which is equal to one for a scale-invariant spectrum, is measured about $n_s=0.9645$ \cite{Planck1}. In contrast with scalar perturbation, Planck does not give an exact value for tensor-to-scalar ratio $r$, it just specifies an upper bound for this parameter as $r<0.10$ \cite{Planck1}.\\

The paper is organized as following: The general form of non-canonical scalar field equations, slow-rolling parameters, and the perturbation parameters are derived in Sec.II. In Sec.III, the problem is studied by using the first formalism, and appropriate values for the free parameters are presented. The second formalism is applied in Sec.IV, where as well as constraining the parameters, the predicted potential, and attractor behavior of the model are considered.\\

\section{Non-canonical Scalar field}\label{Sec2}
To begin, we introduced the Lagrangian which is read as
\begin{equation}\label{1}
\mathcal{S} = \int d^4x \sqrt{-g} \left( {M_p^2 \over 2}\; R + \mathcal{L}_N \right)
\end{equation}
where $R$ is Ricci scalar constructed from the metric $g_{\mu\nu}$, and $M_p$ is Planck mass. $\mathcal{L}_N$ is the Lagrangian of non-canonical scalar field which is defined as $\mathcal{L}_N = F(X) - V(\phi)$. The kinetic term of noncanonical scalar field is expressed by an arbitrary function $F(X)$, in which $X=-(g^{\mu\nu}\nabla_\mu \phi \nabla_\nu \phi)/2$. Potential of scalar field is denoted by $V(\phi)$. \\
The observation shows that the Universe is spatially flat. The common metric to describe such a Universe is spatially flat FLRW metric. The dynamical equations are derived by taking variation of action with respect to independent variable as $g_{\mu\nu}$ and $\phi$. Substituting the metric in the field equations, leads to well-known Friedmann equations as
\begin{equation}\label{2}
H^2 = {1 \over 3M_p^2}\; \rho, \qquad \qquad \dot{H} = {-1 \over 2 M_p^2} \; (\rho + p)
\end{equation}
in which $\rho$ and $p$ are respectively the energy density and pressure of noncanonical scalar field, expressed by
\begin{equation}\label{3}
\rho = 2XF_X - F + V(\phi), \qquad \qquad p= F - V(\phi)
\end{equation}
Derivative of kinetic function $F(X)$ with respect to $X$ is denoted by $F_X$. And for the acceleration equation we have
\begin{equation}\label{4}
{\ddot{a} \over a} = H^2 + \dot{H} = {-1 \over 3M_p^2} \; \Big( F + XF_X - V(\phi) \Big)
\end{equation}
To have a positive acceleration phase for the Universe, one must have $V(\phi) > (F + XF_X)$.
On the other hand, the equation of motion of non-canonical scalar field is obtained
\begin{equation}\label{5}
\big( 2XF_{XX} + F_X \big) \ddot\phi + 3F_X H \dot\phi + V'(\phi) = 0
\end{equation}
which is another expression of familiar conservation equation $\dot\rho + 3H(\rho+p) = 0$, where $\rho$ and $p$ have introduced in Eq.(\ref{3}).\\
From now on, we take the kinetic term as a power-law function of $X$, namely $F(X) = F_0 X^n$. The constant $n$ is dimensionless, and $F_0$ is a constant which its dimension is fitted in a way to give ${\rm [M^4]}$ for kinetic energy density $F(X)$. It could be easily checked that, the case $n=1$ and $F_0=1$ comes to usual canonical scalar field model. \\

\subsection{Non-canonical Inflation}\label{Sec3}
Inflation is an era of the Universe evolution where it stays in a positive accelerated phase and undergoes an extreme expansion. It is suppose that, in this era, the Universe is dominated by an isotropic and homogeneous scalar field which causes quasi-de Sitter expansion. During this work, we assume that, inflation is happens due to a non-canonical inflation, and the general form of parameters are derived.\\

\subsubsection{Slow-Roll Approximation}\label{Sec3-1}
In order to have a quasi-de Sitter expansion, the rate of the Hubble parameter during a Hubble time should be much smaller than unity, in other word $\dot{H}/H^2 \ll 1$ \cite{weinberg}. The same situation is assumed for $\dot\phi$, which states that the rate of time derivative of scalar field during a Hubble time should be much smaller than unity, $|\ddot\phi| / H|\dot\phi| \ll 1$ \cite{weinberg}. These two conditions are known as slow-roll approximations. The first condition allows us to ignore the kinetic energy density of scalar field against the potential part in the Friedmann equation, and the second condition lets one ignore the term $\ddot\phi$ against the term $H\dot\phi$ in the wave equation. Corresponds to each slow-roll approximation there is an slow-roll parameter, given by \cite{weinberg}
\begin{equation}\label{6}
\epsilon_H = -{\dot{H} \over H^2}, \qquad \qquad \eta_H = - {|\ddot\phi| \over H|\dot\phi|}
\end{equation}
Smallness of these two parameter during inflation shows that the scalar field slowly rolls down its potential and let enough amount of inflation happen. \\

\subsubsection{Perturbation}\label{Sec3-1}
Inflationary models predict three kind of perturbations as scalar, vector and tensor perturbation.
one of the main metric perturbation is scalar perturbation. Scalar fluctuations become seeds for cosmic microwave background (CMB) anisotropies or for large scale structure (LSS) formation. Therefore, by measuring the spectra of the CMB anisotropies and density distribution, the corresponding primordial perturbation could be determined. First, let have a brief look at the scalar perturbation.\\
Consider only an arbitrary scalar perturbation to the background FLRW metric, which is expressed by
\cite{Perturbation,Perturbation-a,Perturbation-b, Unnikrishnan,Riotto}
\begin{eqnarray}\label{11}
ds^2 & = & -(1+2A)dt^2 - 2a^2(t)\nabla_i B dx^i dt \nonumber \\
 & & + a^2(t)\Big[(1-2\psi)\delta_{ij} + 2 \nabla_i\nabla_j E  \Big]dx^i dx^j .
\end{eqnarray}
$\delta_{ij}$ is background spatial metric, and $\nabla_i$ is the covariant derivative with respect to this metric. The intrinsic curvature of the spatial hypersurface is expressed by the perturbation parameter $\psi$ as
$^{3}R = {4 \nabla^2 \psi / a^2}$ where $\psi$ is usually named the curvature perturbation \cite{Riotto}. \\
Inserting this metric in the main field equations, leads one to the scalar perturbation, which are expressed in \cite{Perturbation,Perturbation-a,Perturbation-b, Unnikrishnan,Riotto,Bardeen, Hwang}. In this order, we follow \cite{Unnikrishnan}, which has considered the perturbation of generalized gravity, including our model. After some routine algebraic calculation, the results for scalar perturbation show that, the amplitude of scalar perturbation given by \cite{Unnikrishnan}
\begin{equation}\label{12}
\mathcal{P}_s =\left( {H^2 \over 2\pi (c_s (\rho+p))^{1/2}}\right)^{2}
\end{equation}
where $c_A$ is sound speed and for our model is a constant and equal to $c_A^2=(2n-1)^{-1}$ (reader could refer to \cite{Unnikrishnan} for more detail). \\
Besides scalar fluctuation, the inflationary scenario predicts tensor fluctuation, which is known as a gravitational wave, too. The produced tensor fluctuations induce a curved polarization in the CMB radiation and increase the overall amplitude of their anisotropies at a large scale. The physics of the early Universe could be specified by fitting the analytical results the of CMB and density spectra to corresponding observational data. At first, it was thought that the possible effects of primordial gravitational waves are not important and might be ignored. However, a few years ago, it was found out that the tensor fluctuations have an important role, and they should be more attended for determining best-fit values of the cosmological parameters from the CMB and LSS spectra \cite{19, 20, 21}. Contribution of tensor perturbation in metric is expressed as
\begin{equation}\label{13}
ds^2 = -dt^2 + a^2(t)\Big( \delta_{ij} + h_{ij} \Big)dx^i dx^j
\end{equation}
Inseting it into field equations comes to tensor perturbation equations. In contrast with scalar and vector perturbation, energy-momentum perturbation has no role in tensor perturbation equation. After doing some algebraic analysis, the amplitude of tensor perturbation is obtained as \cite{Unnikrishnan}
\begin{equation}\label{14}
\mathcal{P}_T = {8 \over M_p^2} \left({H \over 2\pi}\right)^{2}
\end{equation}
The imprint of tensor fluctuation on the CMB bring this idea to indirectly determine its contribution to power spectra by measuring CMB polarization \cite{20}. Such a contribution could be expressed by the $r$ quantity, which is known as tensor-to-scalar ratio and represents the relative amplitude of tensor-to-scalar
fluctuation, $r=\mathcal{P}_T / \mathcal{P}_s$. Therefore, constraining $r$ is one of the main goals of the modern CMB survey. According to the current accuracy of observations, it is only possible to place a constant upper bound on the allowed range of $r$ \cite{22,22a,22b,22c,22d,22e}. Recent data from nine years of results of WMAP9 and South Pole Telescope (SPT) give the latest constraints of $r<0.13$, and $r<0.11$ at $95\%$ confidence level (C.L.) \cite{23, 24, 25, 26}. Combining Planck's temperature anisotropy measurements with the WMAP large-angle polarization to constrain inflation, gives an upper limit $r<0.11$ in $95\%$ C.L. \cite{DataPot,26}. The latest data about the quantity comes from Planck collaboration on February 2015. Planck full mission data for $\Lambda$CDM+r model resulted in a new constraint on the quantity $r$ as $r<0.10$ (Planck TT,TE,EE+lowP), $<0.11$ (Planck TT+lowP+lensing) at $95\%$ C.L, which indicates an slightly improvement in comparison with the previous result of Planck-2013 \cite{Planck1,Planck2}. \\

\section{Chaotic Inflation}\label{Sec4}
The general form of dynamical equations and main parameters of the model have been derived, which cames to some crud results. In order to have more clear insight, an specific kind of potential is necessary. In this section, we are going to consider the model for a familiar kind of potential, which has received lots of attention, namely power-law potential.\\
Using slow-roll approximations, the dynamical equation of the model are rewritten as
\begin{equation}\label{7}
H^2 = {1 \over 3M_p^2} V(\phi) , \qquad \qquad \dot{H} = -{1 \over M_p^2} \; \Big( n F_0 X^n \Big)
\end{equation}
and the acceleration equation is expressed as $\ddot{a}/a=V(\phi)/3M_p^2$ which shows a positive acceleration, a desirable situation for inflation. Also, for the wave equation, there is
\begin{equation}\label{8}
3 H \dot\phi(nF_0X^{n-1}) + V'(\phi) = 0.
\end{equation}
By utilizing the definition of slow-roll parameters, and the reorganized form of dynamical equations, the slow-roll parameters could be obtained as
\begin{eqnarray}\label{9}
\epsilon_V & = & {3nF_0 \over V(\phi)}\;\left[ {-M_p V'(\phi) \over  nF_0 \sqrt{6V(\phi)} } \right]^{2n \over 2n-1}, \nonumber \\
\eta_V & = & {M_p^2V''(\phi) \over nF_0 V(\phi)} \left[ {nF_0 \sqrt{6V(\phi)} \over -M_p V'(\phi)} \right]^{2(n-1) \over 2n-1}
\end{eqnarray}
This form of slow-roll parameters are known as potential slow-roll parameters, which their smallness display the flatness of potential during inflation. Inflation period lasts until the slow-roll parameter $\epsilon_V$ arrives at one, which corresponds to $\ddot{a}=0$. Amount of expansion during this era is measured by number of e-folds parameter, indicated by $N$ and defined as
\begin{eqnarray}\label{10}
N & \equiv & \ln\left( {a_e \over a_i} \right) = \int_{t_i}^{t_e} H dt = \int_{\phi_i}^{\phi_e} {H \over \dot\phi} d\phi \nonumber \\
 & = & \left({-n F_0 \over 6^{n-1} M_p^{2n}}\right)^{1 \over 2n-1} \int_{\phi_i}^{\phi_e} {\sqrt{V(\phi)} \left(\sqrt{V(\phi)} \over V'(\phi) \right)^{1 \over 2n-1}}\quad
\end{eqnarray}
It is expressed that, to overcome on standard cosmology problems, there should be about $55-65$ number of e-folds.\\

Let's turn our attention to power-law potential $V(\phi)=\lambda \phi^\alpha$. Since the general form of main equations have been acquired in the previous section, we ignored to repeat it, and only express the final results. Substituting this potential in slow-roll parameters (\ref{9}), one arrives at
\begin{equation}\label{15}
\epsilon_V = {3 nF_0 \mathcal{F}^n \over \lambda}  \; \phi^{-z \over \bar{n}}, \qquad \eta_V = {\alpha(\alpha-1)M_p^2 \over nF_0 \mathcal{F}^{n-1}} \; \phi^{-z \over \bar{n}}
\end{equation}
in which
\begin{equation}\nonumber
\mathcal{F} \equiv   \left[ \; {\alpha^2 M_p^2 \lambda \over 6 n^2 F_0^2} \; \right]^{1 \over \bar{n}}, \qquad z \equiv \alpha n + 2n - \alpha, \qquad \bar{n} \equiv 2n-1
\end{equation}
At the end of inflation era, the acceleration parameter vanishes, and the slow-roll parameter $\epsilon_V$ reaches to one. Therefore, at this time, the scalar field could be read from Eq.(\ref{15}) as
\begin{equation}\label{16}
\phi_e^{z  \over \bar{n}} = {3 nF_0 \mathcal{F}^n \over \lambda}
\end{equation}
Using the number of e-folds equation (\ref{10}), one could easily derived the scalar field at the beginning of inflation
\begin{equation}\label{17}
\phi_i^{z  \over \bar{n}} = \mathcal{A} \phi_e^{z  \over \bar{n}}
\end{equation}
in which $\mathcal{A}$ is a constant parameter, defined by $\mathcal{A} \equiv 1 + 2 z N / \alpha \bar{n}$. \\
As it was mentioned, at the beginning of inflation, in order to have a quasi-de Sitter expansion, the slow-roll parameters should be much smaller than unity. Therefore, the constants should be in a way to satisfy this condition. \\
Using the definition of the potential in perturbation amplitude (\ref{12}) and (\ref{14}), the scalar and tensor perturbation amplitudes respectively are obtained as
\begin{equation}\label{r1}
\mathcal{P}_s = { \sqrt{\bar{n}} \; \lambda^2 \over 72\pi^2  M_p^4 \; nF_0 \;\mathcal{F}^n} \; \phi^\alpha \; \phi^{z \over \bar{n}} , \quad \mathcal{P}_T = {2 \over 3\pi^2 M_p^4} \lambda \phi^\alpha \quad \Rightarrow \quad r = {16 \over \sqrt{2n-1}} \; \epsilon_V
\end{equation}
The tensor perturbation is observed indirectly, by using the parameter $r$ which is the ratio of tensor perturbation amplitude to scalar perturbation amplitude, $r=\mathcal{P}_T / \mathcal{P}_s$. These perturbations are used to constraint another free parameter of the model, however first we need to compute them at initial of inflation, namely for $\phi=\phi_i$.\\
On the other hand, the scalar and tensor spectra indices are expressed in term of these slow-roll parameters
\begin{eqnarray}\label{nst}
n_s - 1 & = &  {d\ln(\mathcal{P}_s) \over d\ln(k)} = {2n \over 2n-1}\; \eta_V - {2(5n-2) \over 2n-1}\; \epsilon_V, \nonumber \\
n_T & = & {d\ln(\mathcal{P}_s) \over d\ln(k)} = -2\epsilon_V
\end{eqnarray}

\begin{figure}[ht]
\centering
\subfigure[$n_s-n$]{\includegraphics[width=5.2cm]{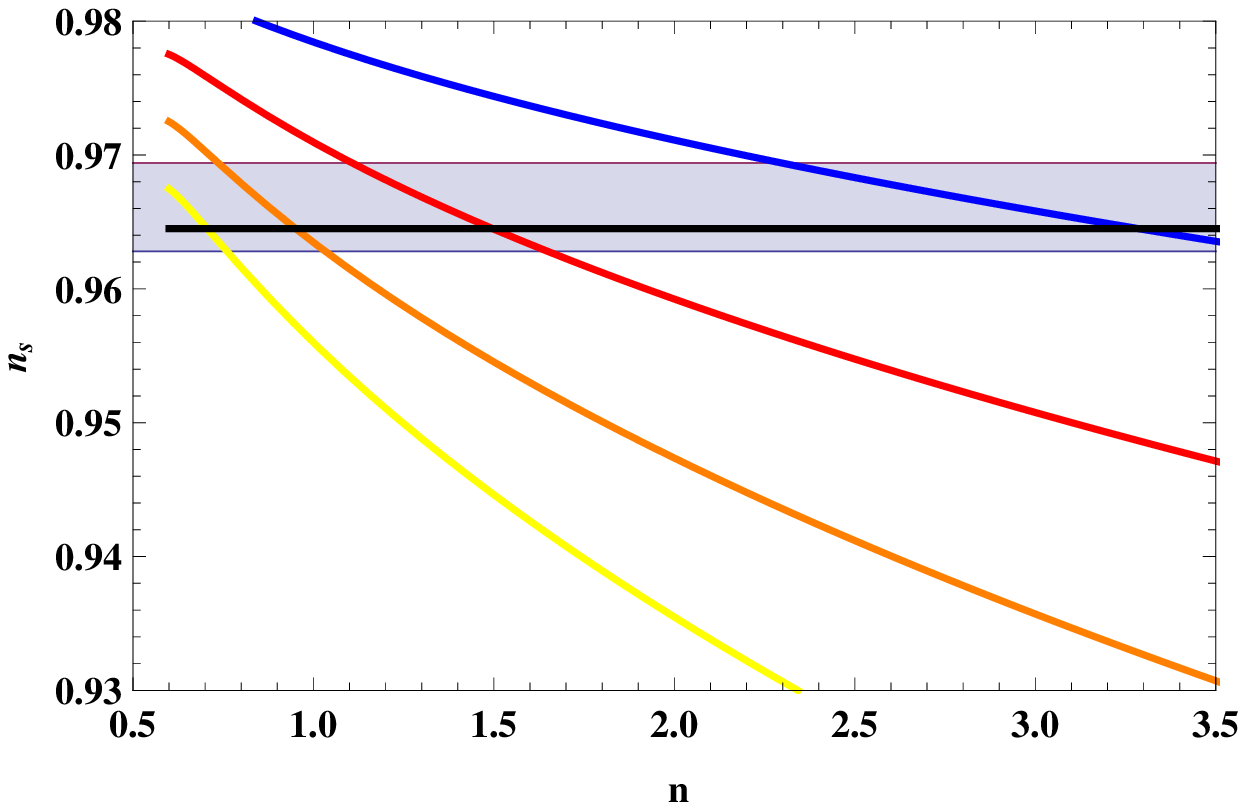}\label{F1a} }
\hspace*{5mm}
\subfigure[$\alpha-n$]{\includegraphics[width=5cm]{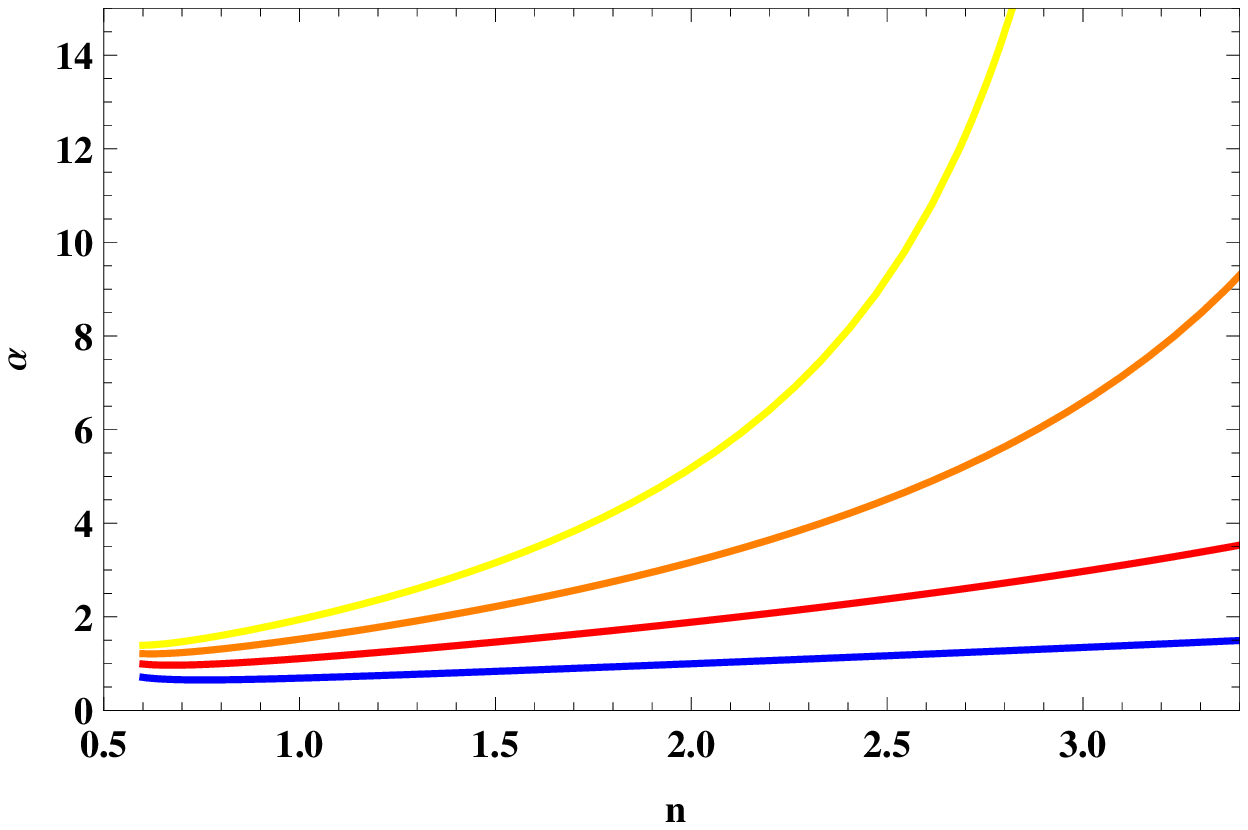}\label{F1b} }
   \caption{a)The scalar spectra index versus $n$; b)The potential power parameter $\alpha$ is depicted versus $n$; for four different values of tensor-to-scalar ratio as $r=0.05$(blue line), $r=0.09$(red line), $r=0.11$(orange line), and $r=0.14$(yellow line). }\label{F01}
\end{figure}

\begin{table}
  \centering
  \begin{tabular}{c|cccc}
     \toprule[1.5pt] \\[-3mm]
     $r$      & \qquad $0.05$ \qquad & \qquad $0.08$ \qquad & \qquad $0.11$ \qquad & \qquad $0.14$  \qquad \\[0.5mm]
     \midrule[1pt] \\[-3mm]
     $n$      & \qquad $3.28$ \qquad & \qquad $1.50$ \qquad & \qquad $0.95$ \qquad & \qquad $0.70$ \qquad \\[2mm]
     $\alpha$ & \qquad $1.45$ \qquad & \qquad $1.46$ \qquad & \qquad $1.46$ \qquad & \qquad $1.47$ \qquad \\[2mm]
     $n_T$    & \qquad $t$ \qquad & \qquad $t$ \qquad & \qquad $t$ \qquad & \qquad $t$ \qquad \\[0.5mm]
     \bottomrule[1.5pt]
   \end{tabular}
  \caption{The final results for the parameters $n$ and $\alpha$.}\label{T01}
\end{table}


\begin{figure}
  \centering
  \includegraphics[width=6cm]{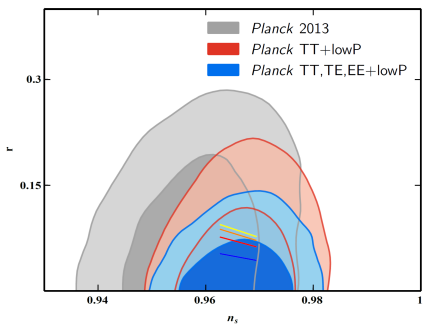}\\
  \caption{The figure shows the tensor-to-scalar ratio versus the scalar spectra index.}\label{F02}
\end{figure}

Utilizing the scalar spectra index (\ref{nst}), and tensor-to-scalar ratio equations (\ref{r1}), one could determine the behavior of the parameter $n$ in term of $n_s$ and $r$. Based on latest Planck data, there are a range for scalar spectra index, and an upper bound for tensor-to-scalar ratio \cite{Planck1}:\\

\begin{tabular}{lccc}
$\ast$ TT+lowP         & $\rightarrow$ & $n_s=0.9655 \pm 0.0062$, \qquad & $r<0.10$, \\
$\ast$ TT+lowP+lensing & $\rightarrow$ & $n_s=0.9677 \pm 0.0060$, \qquad & $r<0.11$,  \\
$\ast$ TT+lowP+BAO     & $\rightarrow$ & $n_s=0.9673 \pm 0.0045$, \qquad & $r<0.11$,  \\
$\ast$ TT,TE,EE+lowP   & $\rightarrow$ & $n_s=0.9645 \pm 0.0049$, \qquad & $r<0.10$.
\end{tabular}

Note that, as we concentrate on Planck-2015 data about the quantity $r$, we realize that the previous mentioned constraint could rise in some cases. For instance, according to \cite{Planck1}, for $\Lambda CDM+r+d\ln n_s / d\ln k$ model, there is $r<0.176$ (Planck TT+lowP+lensing).\\
In order to specify a proper value for the constant parameter $n$, the parameter is depicted versus $n_s$ for four different values of $r$ in Fig.\ref{F1a}. The shadow covers the area which stay between $0.9628<n_s<0.9694$, the common interval for scalar spectra index resulted from Planck-2015 data. The curves cross the shadow line, which in turn indicates interval values of $n$ which are in perfect match with observational data. Fig.\ref{F1b} displays the parameter $\alpha$ in term of $n$ for four different values of tensor-to-scalar ratio. In return to the interval values of $n$, that comes from Fig.\ref{F1a}, one could demonstrate appropriate values for the potential power $\alpha$. The final result has been prepared in the following Table.\ref{T01}. The existence of a modified kinetic term for scalar field is clear from the result, in which the power of kinetic term runs from $0.70$ to $3.28$. In addition, the appropriate potential, to be in agreement with observational data, has a power of $\alpha \simeq 1.46$. \\



\section{Hamilton-Jacobi Approach}
The Hamilton-Jacobi formalism is utilized as a strong tool to provide numerous inflationary models with exactly
known analytic solutions for the background expansion. One of the most interesting advantage of the formalism is that almost the whole parameters of the model could be derived that is explained in the following lines \cite{Baumann,HJ,HJ-a,HJ-b,HJ-c}. Suppose the Hubble parameter could be defined as a function of scalar field $H:=H(\phi)$. Consequently, the time derivative of the Hubble parameter is reexpressed ass $\dot{H}=\dot\phi H'$, where denotes derivative with respect to scalar field. Then, the second Friedmann equation is read as following

\begin{equation}\label{phidot}
\dot{\phi}^{\bar{n}} = -{2^n M_p^2 \over nF_0} \; H'
\end{equation}
where the prime is derivation with respect to scalar field. Taking integrate of the above relation produces a scalar field in term of scalar field, and in turn the hubble parameter could be displayed in term of time as well. Having acknowledge about the Hubble parameter as a function of time, comes to an expression for the universe scale factor. Eventually, The model prediction about the potential is achievable by substituting Eq.(\ref{phidot}) in the Friedmann equation (\ref{2})

\begin{equation}\label{pot}
V(\phi) = 3M_p^2 H^2(\phi) - {\bar{n} F_0 \over 2^n}\; \left( -{2^n M_p^2 \over nF_0} \; H' \right)^{2n \over \bar{n}}
\end{equation}
It could be named as the most important consequence of the formalism, so that for any specific function of $H(\phi)$, the model produces a potential $V (\phi)$ which admits the given $H(\phi)$ as an exact inflationary solution \cite{Baumann,Campo}. Eqs.(\ref{phidot}) and (\ref{pot}) are known as Hamilton-Jacobi equations. \\
The slow-roll parameters of the model are expressed in term of Hubble parameter and its derivatives, in which for the tow well known parameters, we have

\begin{eqnarray}
\epsilon_H & = & - \left( - {2^n M_p^2 \over nF_0} \right)^{1 \over \bar{n}} \; {H'\;^{2n \over \bar{n}} \over H^2} \label{eH} \\
\eta_H     & = & {-1 \over \bar{n}} \left( - {2^n M_p^2 \over nF_0} \right)^{1 \over \bar{n}} \; {H'\;^{2(1-n) \over \bar{n}} H'' \over H} \label{etH}
\end{eqnarray}
Bringing out the same argument as we had in the previous case, one could read the scalar field at the end of inflation. The initial value of scalar field could be derived from number of e-folds relation, so that we have
\begin{equation}
N= \int_{\phi_i}^{\phi_e} {H \over \dot\phi} \; d\phi = \int_{\phi_i}^{\phi_e} \left( - {nF_0 \over 2^n M_p^2}\right)^{1 \over \bar{n}} \; {H \over H'\;^{1 \over \bar{n}}} \; d\phi
\end{equation}
The general form of amplitude of scalar and tensor quantum perturbations were introduced respectively in (\ref{8}) and (\ref{10}). It was explained that how the observational quantities $n_s$ and $n_T$ could be resulted from the corresponding perturbation amplitude. Then, the scalar and tensor spectra indices are given by

\begin{equation}
n_s - 1 = 2n\;\eta_H - 4\epsilon_H; \qquad n_T = -2\epsilon_H.
\end{equation}
All required equation for considering inflation era using Hamilton-Jacobi approach, were introduced so far. The next step for going ahead, and studying the situation in more detail is to pick out a specific function for the Hubble parameter in term of scalar field. Then, it is assumed that the hubble parameter could be expressed as a power-law function of scalar field, so that $H(\phi)=H_0 \phi^\beta$. \\
Substituting the expression of the Hubble parameter in the slow-roll equations (\ref{eH}) and (\ref{etH}), comes to the following consequences

\begin{eqnarray}
\epsilon_H & = &  - \left( - {2^n M_p^2 \over nF_0} \right)^{1 \over \bar{n}}
{\beta^{2n \over \bar{n}} \over H_0^{2(n-1) \over \bar{n}}}\; \phi^{\sigma \over \bar{n}}  \\
\eta_H     & = &  -{(\beta-1)\beta^{1 \over \bar{n}} \over \bar{n}} \left( - {2^n M_p^2 \over nF_0} \right)^{1 \over \bar{n}} H_0^{2(1-n) \over \bar{n}}\; \phi^{\sigma \over \bar{n}}
\end{eqnarray}
where the constant parameter $\sigma$ is defined as $\sigma = 2\beta - 2n \beta - 2n$. The inflation ends when the slow-roll parameter $\epsilon_H$ reach unity, which in turn describe $\ddot{a}=0$, and the final scalar field could be read easily. As it was mentioned, the scalar field at the beginning of inflation is acquired from number of e-folds relation as well. Final and initial scalar field respectively is resulted as

\begin{eqnarray}
\phi_e^{\sigma \over \bar{n}} & = & - \left( - {nF_0 \over 2^n M_p^2} \right)^{1 \over \bar{n}}
{H_0^{2(n-1) \over \bar{n}} \over \beta^{2n \over \bar{n}}} \\
\phi_i^{\sigma \over \bar{n}} & = & {1 \over \mathcal{B}} \phi_e^{\sigma \over \bar{n}}
\end{eqnarray}
where the redefined constant $\mathcal{B}$ is given by $\mathcal{B}=1-\sigma N/\beta$.\\
Consequently, the perturbation quantities could be derived at the beginning of inflation by inserting the initial value of scalar field in the corresponding relations. Therefore, the spectra indices are described as

\begin{equation}
n_s-1 = {4\beta-6n\beta-2n \over \bar{n}\beta\mathcal{B}}; \qquad n_T=- {2 \over \mathcal{B}},
\end{equation}
and the tensor-to-scalar ratio, which indirectly indicates the presence of gravitational waves, comes to the following expression
\begin{equation} \label{r2}
r = {16c_A^2 \over \mathcal{B}}   \qquad {\rm or} \qquad   r =  \; {16c_A^2\bar{n}\beta \over 4\beta-6n\beta-2n} \; (n_s - 1).
\end{equation}

\begin{figure}[h]
   \centering
   \includegraphics[width=5cm]{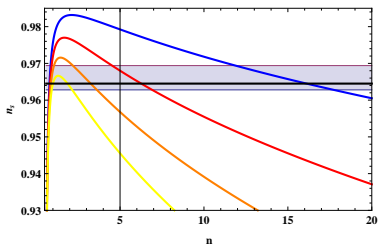}
   \caption{The scalar spectra index versus $n$ for four different values of tensor-to-scalar ratio, as $r=0.05$(blue line), $r=0.09$(red line), $r=0.11$(orange line), and $r=0.14$(yellow line). }\label{F03}
\end{figure}

Utilizing the relations for $n_s$, $r$, and following the same process as previous section, the behavior of the constant parameter $n$ could be specified easily in term of $n_s$ and $r$. Fig.\ref{F03} portrays the behavior versus scalar spectra index for four different values of $r$. In comparison with the previous method, there are two separate interval for $n$ which could be consistence with observational data. \\

\begin{figure}[h]
\centering
\subfigure[$r=0.08$]{\includegraphics[width=4cm]{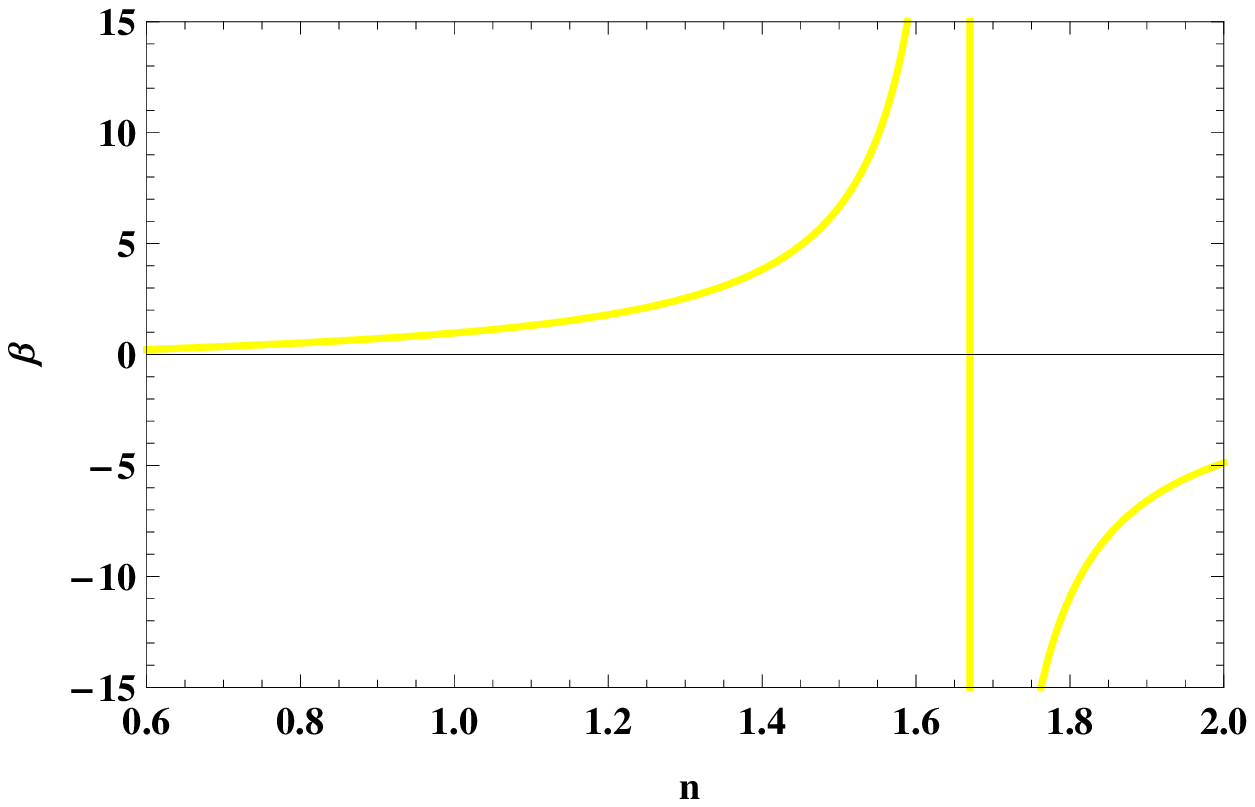}\label{F4a} }
\hspace*{5mm}
\subfigure[$r=0.08$]{\includegraphics[width=4cm]{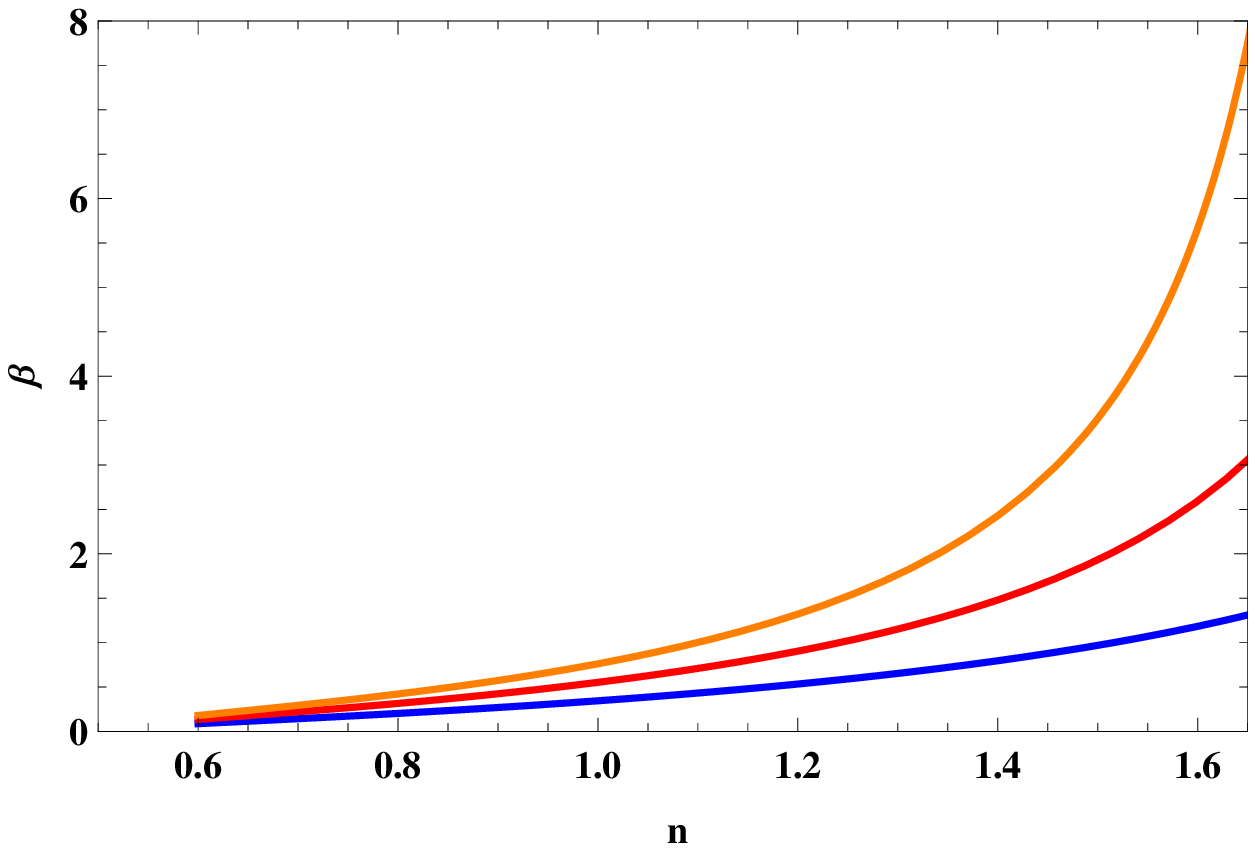}\label{F4b} }
\hspace*{5mm}
\subfigure[$r=0.10$]{\includegraphics[width=4cm]{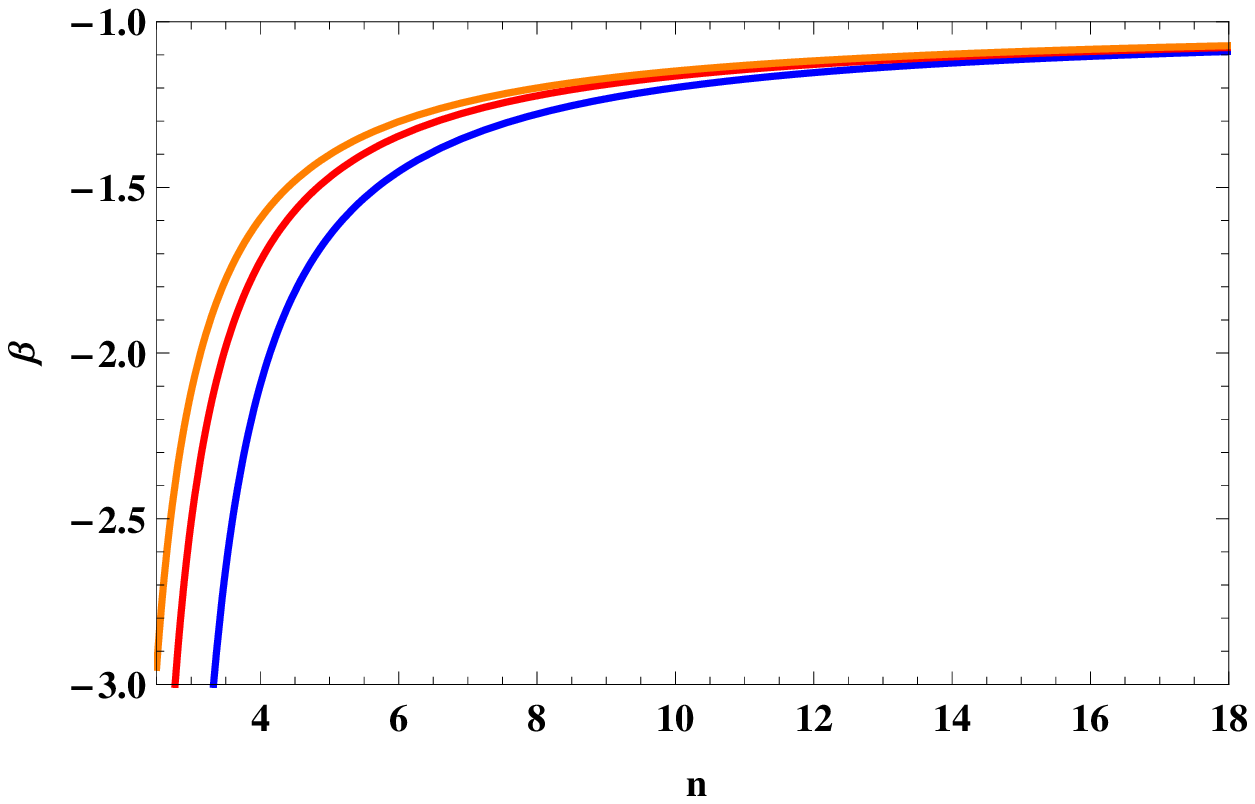}\label{F4c} }
\caption{The Hubble power parameter $\beta$ is depicted versus $n$ for four different values of tensor-to-scalar ratio, as $r=0.05$(blue line), $r=0.08$(red line), $r=0.11$(orange line), and $r=0.14$(yellow line).}\label{F04}
\end{figure}

\noindent Definitely, for these interval of $n$, there are two interval for the Hubble power parameter $\beta$. The behavior of the parameter versus $n$ is illustrated in Fig\ref{F04} for four different values of $r$. Fig.\ref{F4a} is plotted for $r=0.05$. Fig.\ref{F4b} and Fig.\ref{F4c} depicts the behavior of $\beta$ versus $n$ to identify proper values of $\beta$ respectively for the first and second interval of $n$. Ultimate result has been expressed in the Table.\ref{T02}. \\

\begin{table}[h]
  \centering
  \begin{tabular}{c|cccc}
     \toprule[1.5pt] \\[-3mm]
     $r$      & \qquad $0.05$ \qquad & \qquad $0.08$ \qquad & \qquad $0.11$ \qquad & \qquad $0.14$  \qquad \\[0.5mm]
     \midrule[1pt] \\[-3mm]
     $n_1$      & \qquad $0.80$ \qquad & \qquad $0.83$ \qquad & \qquad $0.88$ \qquad & \qquad $1.00$ \qquad \\[2mm]
     $\beta_1$ & \qquad $0.20$ \qquad & \qquad $0.34$ \qquad & \qquad $0.54$ \qquad & \qquad $0.97$ \qquad \\[2mm]
     \hline \\[-2mm]
     $n_2$      & \qquad $16.12$ \qquad & \qquad $6.26$ \qquad & \qquad $3.25$ \qquad & \qquad $1.86$ \qquad \\[2mm]
     $\beta_2$ & \qquad $-1.10$ \qquad & \qquad $-1.32$ \qquad & \qquad $-1.92$ \qquad & \qquad $-7.77$ \qquad \\[0.5mm]
     \bottomrule[1.5pt]
   \end{tabular}
  \caption{The final results for the parameters $n$ and $\alpha$.}\label{T02}
\end{table}

Hamilton-Jacobi formalism again indicates the necessity of modified version of kinetic term for scalar field, however, the difference is that the power of kinetic term sounds to be smaller than unity for $n_1$, about $n_1 \simeq 0.8$, and larger than previous case for $n_2$, about about $n_2 \simeq 0.8$. The result for power of Hubble parameter displays that it could be both positive for $n_1$, or negative for $n_2$. The whole result has been presented in Table.\ref{T02}.  \\

\begin{figure}
\centering
\subfigure[$r=0.08$]{\includegraphics[width=6cm]{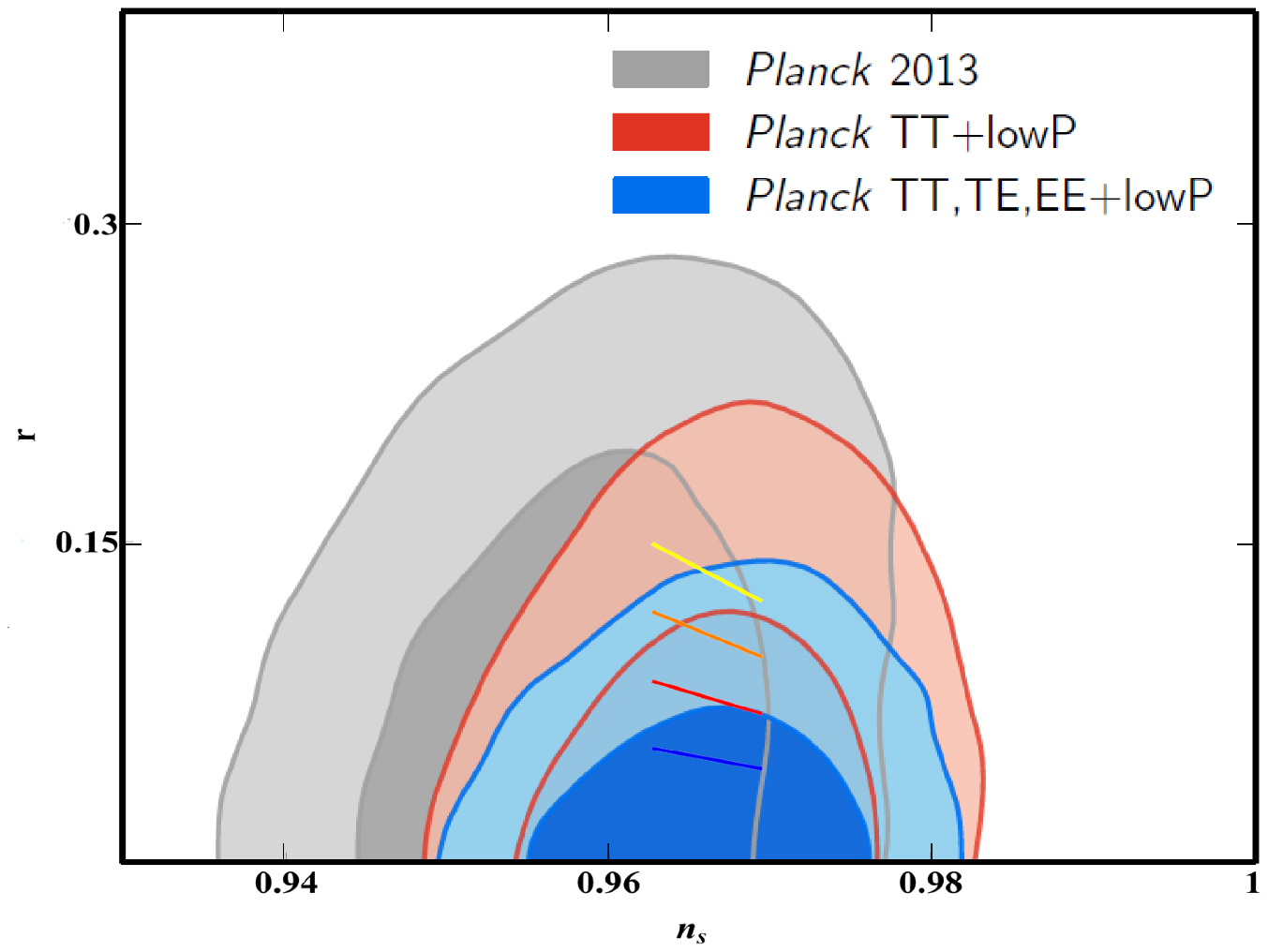}\label{F7a} }
\hspace*{5mm}
\subfigure[$r=0.10$]{\includegraphics[width=6cm]{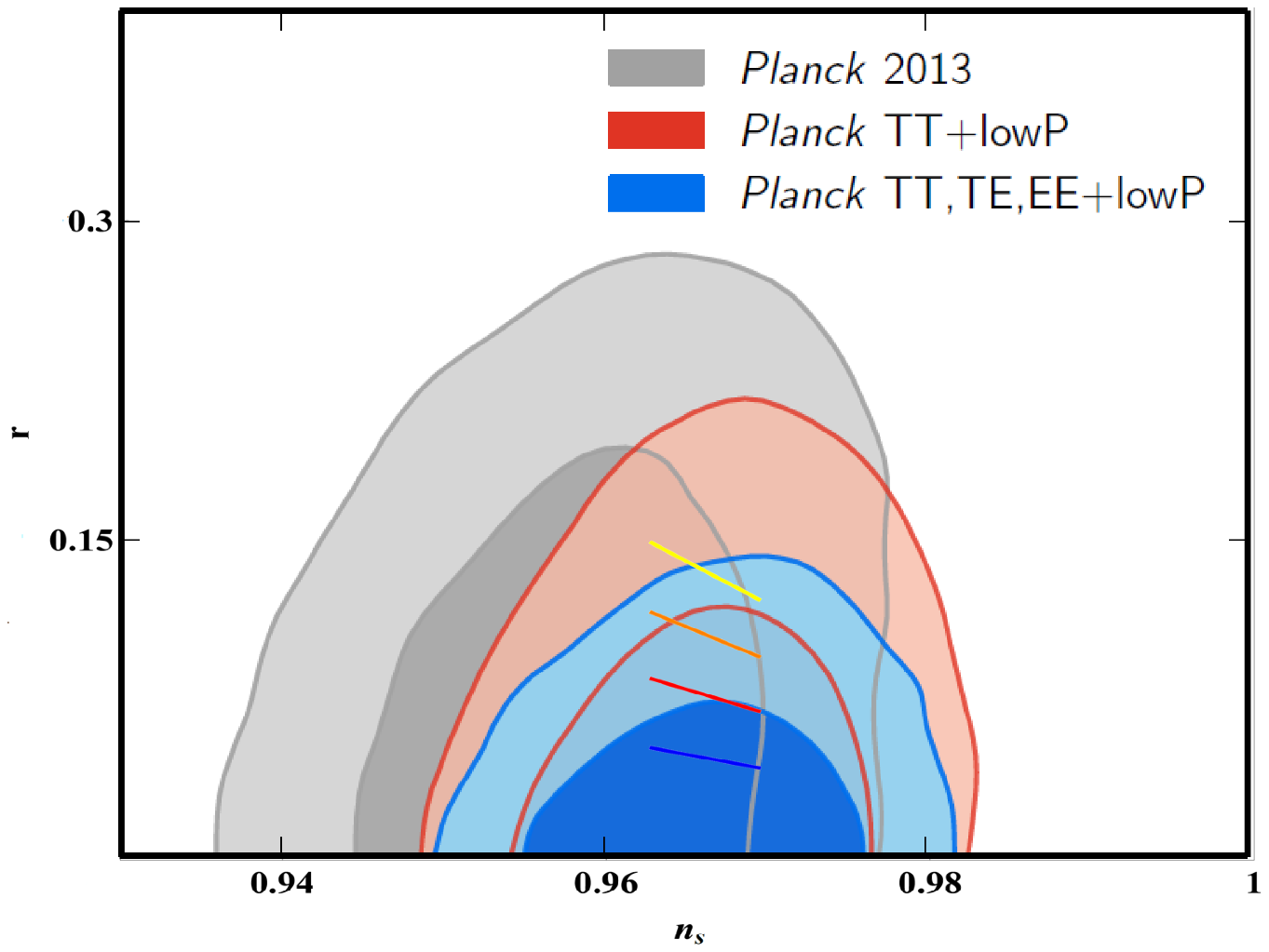}\label{F7b} }
\caption{ The figure shows the tensor-to-scalar ratio versus the scalar spectra index.}\label{F05}
\end{figure}

So far, the constant parameter $n$ and $\beta$ have been determined by using observational data. Then, these results could be applied in Eq.(\ref{r2}), to consider the behavior of $r$ in term of $n_s$, as shown in Fig.\ref{F05} for two separate set of $(n,\beta)$. The results is almost same for both, in which it proves that the free parameters have been chosen properly. \\

\subsection{Potential}
The general form of the potential was introduced in Eq.(\ref{pot}), and it is a polynomial in $\phi$. In order to study the behavior of the potential during the inflationary times, the potential could be drew in term of scalar field using the constant parameter $n$, and $\beta$ from Table.\ref{T02}. It has been depicted for four different values of tensor-to-scalar ration, displayed in Fig.\ref{Pot}. In all cases, the potential reduces with increasing time (or decreasing scalar field), and the only difference stand in the slip of the potential curve. The Potential diagram is convex, and with increasing $r$, it alters and turns to a convex diagram. For $r=0.11$, it looks like a linea potential when it behaves as $V \propto \phi$; and for $r=0.14$, it behaves like the well-known and common potential $V \propto \phi^2$. Note that, During inflationary times, the initial and final values of the scalar field are smaller thank Planck mass in the first three cases; and in the final plot (where $r=0.14$), only the initial values of scalar field becomes larger than Planck mass. Then, it is resulted that the inflation could occur for field values below the Planck mass, almost the same result was obtained in \cite{Maartens}, where the authors study chaotic inflation in brane-world scenario. In addition, as it was expected, the potential energy density is always smaller than Planck energy.  \\

\begin{figure}[ht]
\centering \includegraphics[width=10cm]{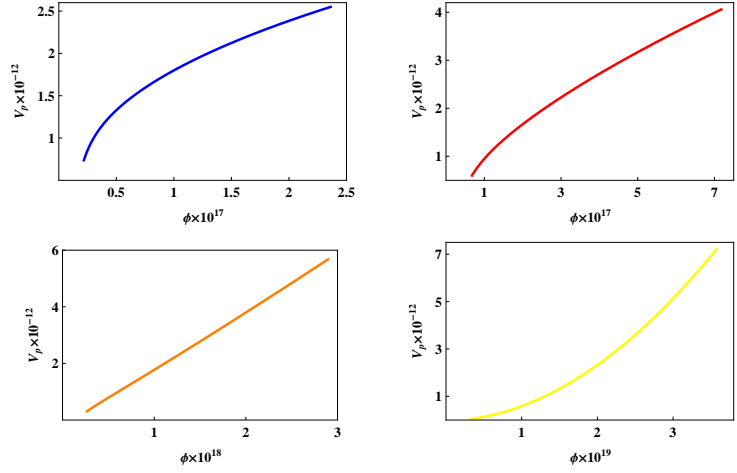}\label{Pot}
\caption{ The potential behavior has been portrayed for different values of tensor-to-scalar ratio as: $r=0.05$(Blue), $r=0.08$(Red), $r=0.11$(Orange), and $r=0.14$(Yellow).}\label{Pot}
\end{figure}

\subsection{Attractive Behavior}
Following \cite{Lyth}, the attractive behavior of the model could be investigated by adding a homogeneous perturbation $\delta H(\phi)$ to a hypothetical solution $H_0(\phi)$ of Hamilton-Jacobi equation (\ref{pot}). If the expression $\delta H(\phi) / H_0(\phi)$ approach to zero by passing time, the attractor condition could be satisfied. Substituting $H(\phi)=H_0(\phi)+\delta H(\phi)$ in the Hamilton-Jacobi equation and linearizing, the perturbation equation is read as

\begin{equation}
{H'}_0^{1 \over \bar{n}} \delta H' = - {3\bar{n} \over 2n} \left( - {nF_0 \over 2^n M_p^2} \right)^{1 \over \bar{n}} \; H_0 \delta H.
\end{equation}
Taking integrate of above equation, one arrives at the following expression for the perturbation $\delta H$

\begin{equation}\label{deltaH}
\delta H = \delta H(\phi_i) \exp\left( -{3\bar{n} \over 2n}  \; N \right).
\end{equation}
Eq.(\ref{deltaH}) strongly points out the fact that by passing time, the perturbation parameter $\delta H$ vanishes, and the model successfully satisfies the attractor condition. \\

\section{Special Case}
The kinetic term of scalar field was taken as a power-law function of $X$, namely $F(X)=F_0 X^n$. For such a kinetic energy, one is faced to a special case when the power of the term is taken as $n=1/2$. In this section, we are going to briefly review the case.\\
Substituting $n=1/2$ in the general dynamical equations obtained in Sec.\ref{Sec2}, comes to following result for Friedmann equations
\begin{eqnarray}
H^2 & = & {1 \over 3M_p^2}\; V(\phi) \label{S-Friedmann1}\\
\dot{H} & = & -{F_0 \over 2M_p^2}\; \sqrt{X}. \label{S-Friedmann2}
\end{eqnarray}
The first surprising conclusion is that the energy density of scalar field is exhibited by potential energy density, and the kinetic term has no contribution. Also, taking integrate of Eq.(\ref{S-Friedmann2}) gives the Hubble parameter in term of scalar field,
\begin{equation}
H(\phi) = {\mp F_0 \over 2\sqrt{2}\; M_p^2}\; \phi + c_1;
\end{equation}
where $c_1$ is a constant of integration, and $\mp$ respectively are related to $\sqrt{X} \propto \dot\phi$ and $\sqrt{X} \propto -\dot\phi$. In addition to the Hubble parameter, the potential energy density could be derived exactly from the equation of motion of scalar field, so that
\begin{equation}
{3F_0 \over 2}\; H\; {\sqrt{2} \over \dot\phi}\; \phi + V'(\phi)=0
\end{equation}
and by taking integrate and utilizing the first Friedmann equation (\ref{S-Friedmann1}), the potential is expressed by
\begin{equation}
V(\phi) = {1 \over 4} \; \left( {\mp 3F_0 \over \sqrt{6}\; M_p}\; \phi + c_2 \right)^2,
\end{equation}
in which $c_2$ is constant of integration, related to $c_1$ by $c_2=2M_p\sqrt{3}\; c_1$. Note that, by re-scaling the scalar field, the potential could be rewritten as a potential of massive scalar field $V(\varphi)=m_\varphi^2 \varphi^2 / 2$, so that $\varphi\equiv \mp \phi + {2\sqrt{2} M_p^2 c_1 \over F_0}$, and $m_\varphi \equiv \sqrt{3 \over 4}\; {F_0 \over M_p}$. \\
As a further argument, it should be mentioned that, the dynamical equations of this special case are automatically in the form of the dynamical equation of usual cases after applying the slow-rolling approximation.

\section{Conclusion}
The inflationary scenario was studies using a non-canonical scalar field instead of usual canonical scalar field. It was supposed that the kinetic term, $F(X)$, is a power-law function of $X$($=\nabla_\mu\phi\nabla^\mu\phi/2$), namely there is $F(X)=F_0X^n$ where $F_0$ is a constant parameter. In the slow-roll inflationary scenario, regardless of whether scalar field in canonical or non-canonical, the time rate of the Hubble parameter during a Hubble time should be smaller than unity, and the same behavior is supposed for time derivative of scalar field; known as slow-roll approximation. The work was implemented by using two formalisms, and the main goal was to constrain the free parameter of the model coming from the latest observational data. \\

Recent observational data gives us a proper insight about some perturbation parameters such as spectra indices and amplitude of perturbations. The latest data is related to Planck, released on 2015. The result exhibits a slightly correction to Planck 2013, and states that the scalar spectra index is about $n_s =  0.9645 \pm ...$, and the amplitude of the perturbation is measured as $\ln \Big( 10^{10}\mathcal{P}_s \Big)=3.094$. The tensor perturbation is still not clear as the scalar perturbation so that we only have an upper bound for tensor-to-scalar ration as $r<0.10$. \\

To start, the first formalism was used and a general form of a power-law potential was proposed to consider the situation in detail. After obtaining the general form of evolution equations, and using slow-roll approximations, the slow-roll parameters $\epsilon$ and $\eta$ were derived for non-canonical scalar field model including a power-law kinetic term. As it was expected, these parameters come back to the standard form of slow-roll parameters for $n=F_0=1$. Utilizing the Planck data, we determined the power of kinetic term and potential, namely $n$ and $\alpha$; presented in Table.\ref{T01}. The power of kinetic energy term is obtained as $n \approx 0.70, 1.5, 3$ which indicates the necessity of a non-canonical term for kinetic energy of scalar field. In addition, the constraint values for power of potential function showed that this parameter should be about $\alpha \approx 1.46$ which are in good agreement with the recent observational results about the form of potential in inflationary era.\\

At next step, the Hamilton-Jacobi formalism was applied by presenting a power-law function of scalar field for the Hubble parameter, $H \propto \phi^\beta$. Repeating the same process resulted in a non-canonical kinetic term for scalar field so that the parameter $n$ changes from $0.80$ to $1.00$; and the power of the Hubble parameter $\beta$ stands between $0.20-0.97$. The constraint values were be utilized to portray the predicted potential of the model. By contrast to inflationary studies using the first formalism, considering the potential behavior points out that the inflation could occur even when the field values stands below the Planck mass. During inflation times, the potential decreases with reduction of scalar field.



\end{document}